\documentclass[9pt,twocolumn,twoside]{osajnl}
\journal{ol} % Choose journal (ao, aop, josaa, josab, ol, pr)

\setboolean{shortarticle}{true}
\newcommand{\revision}[1]{{\color{black} #1}}

\title{Edge mode bifurcations of two-dimensional topological lasers}

\author[1,*]{Sinan G\"{u}ndo\u{g}du}
\author[1,2]{Juzar Thingna}
\author[1,2]{Daniel Leykam}

\affil[1]{Center for Theoretical Physics of Complex Systems, Institute for Basic Science, Daejeon 34126, Korea}

\affil[2]{Basic Science Program, Korea University of Science and Technology, Daejeon 34113, Korea}

\affil[*]{Corresponding author: gu.sinan@gmail.com}

\begin{abstract}
Topological lasers are of growing interest as a way to achieve disorder-robust single mode lasing using arrays of coupled resonators. We study lasing in a two-dimensional coupled resonator lattice exhibiting transitions between trivial and topological phases, which allows us to systematically characterize the lasing modes throughout a topological phase. We show that, unlike conventional topological robustness that requires a sufficiently large bulk band gap, bifurcations in topological edge mode lasing can occur even when the band gap is maximized. We show that linear mode bifurcations from single mode to multi-mode lasing can occur deep within the topological phase, sensitive to both the pump shape and lattice geometry. We suggest ways to suppress these bifurcations and preserve single edge mode lasing.
\end{abstract}

\setboolean{displaycopyright}{true}

\begin{document}

\maketitle

%\section{Introduction}

Topological photonics is a rapidly growing research field based on controlling the flow of light using concepts from condensed matter physics~\cite{Ozawa2019}. A key idea is that energy bands with non-trivial topological invariants give rise to families of special edge modes in band gaps. The edge modes are protected against backscattering due to certain classes of disorder, which can help us to achieve disorder-immune waveguiding and manipulation of light. There is growing interest in topological lasers based on adding gain to topological photonic systems to make their protected edge modes lase~\cite{smirnova2019nonlinear,Ota_2020,Feng2017}. Recent demonstrations of topological lasers have used either photonic crystals~\cite{Bahari2017,Ota2018,Han2019,Zeng2020} or photonic lattices of weakly coupled resonators~\cite{StJean2017,Parto2018,Zhao2018,Harari2018,Bandres2018}. Our focus here is on the latter.

Photonic lattice-based topological lasers can be further divided into two classes: one-dimensional (1D) and two-dimensional (2D). 1D experiments were based on the Su-Schrieffer-Heeger model hosting mid-gap edge modes localized to one of its two sublattices, which spontaneously break the model's sublattice symmetry~\cite{StJean2017,Parto2018,Zhao2018}. When gain is applied by pumping a single sublattice, only the edge modes are amplified, while the bulk modes which are distributed over both sublattices do not lase. As the pump strength is increased the bulk band gap narrows, and eventually closes, accompanied by the emergence of multi-mode lasing of the bulk modes~\cite{Schomerus2013,Weimann2016}.

2D experiments and theory have been largely based on the Harper-Hofstadter model, which describes a lattice with inhomogeneous nearest neighbor hopping phases~\cite{Hafezi2011,Bandres2018}. It hosts a family of edge modes which traverse the bulk band gap, wrap around the entire edge of the lattice, and can be made to lase by localizing the pump to the lattice edge. However, the dynamics can be quite complicated due to competition between multiple edge states with similar gain, leading to peculiar spatial and temporal coherence properties~\cite{Secl2019,Amelio2019}. For example, irregular pulsating dynamics emerge in class B laser models of these lattices~\cite{Longhi2018}. Systematic studies of the laser performance within the topological phase remain limited. For example, should one maximize the size of the topological band gap to achieve the most robust single mode lasing, as suggested in Ref.~\cite{Harari2018}? Or do other non-topological effects play a more important role, such as the shape of the lattice boundary or pump profile?

The aim of this Letter is to address these questions by studying a Haldane-like model of a tunable topological resonator lattice~\cite{Leykam2018,Mittal2019}. This is perhaps the simplest example of a two-dimensional topological lattice, because it exhibits a single bandgap that can be either topologically trivial or non-trivial, depending on the frequency detuning of its two sublattices. We find that even in the linear limit close to the lasing threshold, transitions in the lasing modes can occur deep within the topological phase, unaccompanied by any bulk topological transition. For example, beyond a critical sublattice detuning, midgap single mode lasing is replaced by lasing of two modes with frequency detunings $\pm \omega$ with respect to the middle of the band gap. These transitions arise due to the interplay between the pump and edge mode profiles; localizing the pump to the edge can suppress the lasing of bulk modes, but generally further fine-tuning is generally required to enforce lasing in a single edge mode. We validate our linear mode analysis with time domain simulations of a class A laser model, and suggest strategies to enforce single edge mode lasing. 

\begin{figure}
\centering\includegraphics[width=\columnwidth]{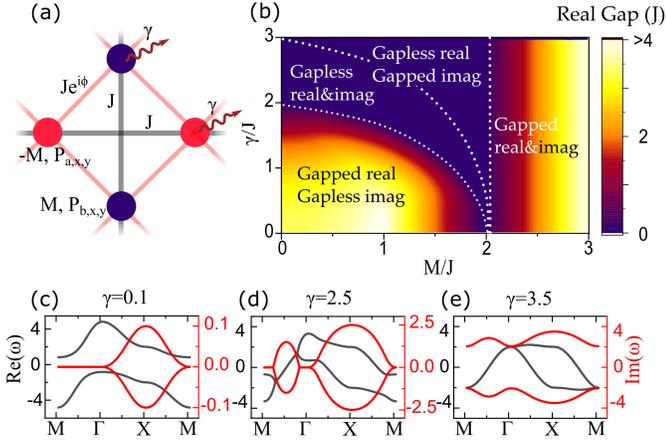}
\caption{(a) Unit cell of the lattice, consisting of two sublattices indicated with blue and red discs.  Each sublattice connect the nearest and the next nearest node with hopping strength $J$. The hopping phase between the two sublattices is $\phi=\pm \pi/4$. The two sublattices have loss $\gamma$, pump rates $P_{a,x,y}$ and $P_{b,x,y}$, and detunings $\pm M$ with respect to a reference frequency. (b) Phase diagram of the Bloch band structure of the linear tight binding Hamiltonian with symmetric gain/loss ($P_{a,x,y} = 2\gamma, P_{b,x,y}=0$), as a function of detuning and $\gamma$. \revision{Color indicates the real part of the gap, i.e. the difference between lowest frequency of the upper band and the highest frequency of the lower band.} (c-e) Bloch band structures for $M=0$ and $\gamma=0.1J$ (c), $2.5J$ (d) and $3.5J$ (e).}
\label{fig:lattice}
\end{figure}

We consider a recently-demonstrated ring resonator lattice with tunable topological properties~\cite{Leykam2018,Mittal2019}. Its unit cell is shown in Fig.~\ref{fig:lattice}(a), consisting of two sublattices ($a,b$) whose resonant frequencies can be independently tuned. Each resonator is coupled to its nearest neighbors (shown with diagonal lines) as well as next-nearest neighbors (horizontal and vertical lines). Coupling is mediated by anti-resonant link rings, resulting in the same nearest and next-nearest neighbor coupling strengths, as well as a direction-dependent hopping phase for the nearest neighbor coupling. Considering a single circulation direction within the resonant rings (anti-clockwise), and assuming negligible backscattering between clockwise and anit-clockwise modes, the linear optical field dynamics in this lattice are well-approximated by the tight binding Hamiltonian
\begin{align}
    \hat H&=\sum_{x,y}{(\hat{H}_a+\hat{H}_b+\hat{H}_{ab}+\hat{H}^\dagger_{ab})},\\
    \hat H_{a}&=\hat{a}^\dagger_{x,y} \left( (M + i P_{a,x,y} - i \gamma )\hat{a}_{x,y} +
    J  \sum_{\pm} \hat{a}_{x,y\pm1} 
    \right), \nonumber \\
    \hat H_{b}&=\hat{b}^\dagger_{x,y} \left( (-M + i P_{b,x,y} - i \gamma )\hat{b}_{x,y} +
    J  \sum_{\pm} \hat{b}_{x\pm1,y} 
    \right), \nonumber \\
    \hat H_{ab}&=J e^{i\pi/4} \left(\hat{a}^\dagger_{x,y}(\hat{b}_{x,y}+\hat{b}_{x+1,y+1})+\hat{b}^\dagger_{x,y}(\hat{a} _{x-1,y}+\hat{a} _{x,y-1})\right), \nonumber
\end{align}
where $\hat{a}^\dagger_{x,y} (\hat{a}_{x,y})$ and $\hat{b}^\dagger_{x,y} (\hat{b}_{x,y})$ are photon creation (annihilation) operators for the $a$ and $b$ sublattice sites in unit cell $(x,y)$, $J$ is the coupling strength, $M$ is the sublattice detuning, $P_{\psi, x,y}$ is the normalized local pump rate, and $\gamma$ accounts for scattering losses. As $\hat{H}$ is non-Hermitian, its eigenvalues $\omega$ are in general complex, with the real and imaginary parts describing the modal frequenies and small signal gain, respectively. We use a class A laser model to describe the saturation of gain at high intensities, such that classical optical field amplitudes evolve according to
\begin{equation}
   i\partial_t \psi_{x,y} = \left( \hat{H} - \frac {i P_{\psi,x,y} |\psi_{x,y}|^2 / I_{\mathrm{sat}}}{1+|\psi_{x,y}|^2 /I_{\mathrm{sat}}} \right) \psi_{x,y}, \quad \psi=a,b\label{eq:classA}
\end{equation}
 where $I_{sat}$ is the saturation intensity, which we normalize to 1 without loss of generality. We will start by analyzing \eqref{eq:classA} in the linear limit, which describes the initial build up of the lasing modes before the onset of gain saturation.

First we consider a bulk pumping configuration similar to that previously used for the Su-Schrieffer-Heeger model~\cite{Schomerus2013,Weimann2016,StJean2017,Parto2018,Zhao2018}, where all sites belonging to one sublattice are pumped and we assign an equal amount of loss to the other sublattice, i.e. setting $P_{a,x,y} = 2 \gamma$ and $P_{b,x,y} = 0$. We compute the bulk spectrum by Fourier transforming the tight binding Hamiltonian into momentum $\boldsymbol{k} = (k_x,k_y)$ space, yielding the Bloch Hamiltonian
\begin{align}
&\hat{H}(\boldsymbol{k}) = \left(\begin{array}{cc} H_a & H_{ab} \\ H_{ab}^* & H_b \end{array} \right),\\
&H_a = 2 J \cos k_y + M + i \gamma, \nonumber\\
&H_b = 2 J \cos k_x - M - i \gamma, \nonumber\\
&H_{ab} = J e^{i \pi/4} ( 1 +e^{i (k_x+k_y)}) + J e^{-i\pi/4} (e^{i k_x} + e^{i k_y}).\nonumber
\end{align}
The eigenvalues of $\hat{H}(\boldsymbol{k})$ \revision{are $\omega_{\pm}(\boldsymbol{k})$ and} form two bands, which are classified in Fig.~\ref{fig:lattice}(b) as a function of the gain/loss contrast $\gamma$ and the sublattice detuning $M$. Increasing the gain/loss contrast induces a narrowing of the \revision{real part of} bulk band gap, \revision{i.e. $\mathrm{min}_{\boldsymbol{k}} (\mathrm{Re}[\omega_+]) - \mathrm{max}_{\boldsymbol{k}}( \mathrm{Re}[\omega_-])$}. At strong pump rates the coupling $J$ becomes negligible and the modes of each band are strongly localized to a single sublattice, corresponding to a gap in the imaginary part of the eigenvalue spectrum. If the lattice is initially in the topological phase ($M<2J$), at moderate pump rates there is an intermediate phase in which both the real and imaginary parts of the energy eigenvalues are gapless, exhibiting non-Hermitian degeneracies at critical momenta, similar to the Su-Schrieffer-Heeger model~\cite{Schomerus2013,Weimann2016}. 

Representative examples of the band structures in the different phases are shown in Fig.~\ref{fig:lattice}(c-e), revealing that the dominant modes, i.e. those with the highest linear gain, are insensitive to the opening and closing of the band gap and always occur at the X points $\boldsymbol{k} = (0,\pi)$ and $(\pi,0)$, where $H_{ab} = 0$. More generally in two-dimensional topological models, regardless of sublattice polarization of a uniform pump there will always be some bulk Bloch modes which overlap perfectly with the pump and have the highest gain. Thus, additional spatial structuring of the pump profile is necessary to achieve edge mode lasing.

\begin{figure}
\centering\includegraphics[width=\columnwidth]{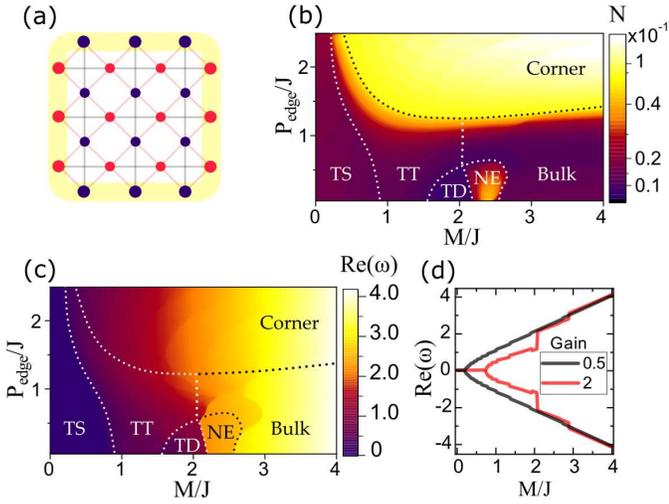}
\caption{(a) Edge-pumped 4x4 $C_4$-symmetric lattice. Only the sites in the yellow highlighted region are pumped. (b,c) Participation number (b) and frequency (c) of the dominant mode with the largest gain as a function of detuning and pump rate, for a 12x12 $C_4$ symmetric lattice under edge pumping. Different lasing regimes are shown; TS: topological edge mode lasing with single dominant mode. TT: two mode topological edge state lasing. TD: topological edge modes starting to delocalize into the bulk. NE: non-topological (trivial) edge modes.  (d) Frequency bifurcations of the the dominant mode as a function of detuning $M$, for pump rates $P_{edge} = 0.5J$ and $2J$.}
\label{fig:fig2}
\end{figure}

Next, we consider the spectrum of moderately-sized finite lattices where only the edge resonators are pumped, which can support either bulk or edge mode lasing, i.e. $P_{a/b,\bar x,\bar y}=P_{edge}$, where $\bar x$, $\bar y$ denote edge sites. Two different edge configurations are of interest: the edges can be terminated always with elements of the same sublattice ($a$ or $b$), or such that the $C_4$ discrete rotational symmetry of the lattice is preserved, corresponding to the horizontal edges being terminated with $a$ sites, and the vertical edges with $b$ sites. In this work, we focus on the latter case shown in Fig.~\ref{fig:fig2}(a), as the single sublattice termination typically exhibits strong competition between bulk and non-topological corner modes.

\revision{We identify the dominant lasing mode as the eigenmode whose imaginary part of its eigenvalue is largest.} To characterize this dominant lasing mode as a function of the detuning and pump rate in Figs~\ref{fig:fig2}(b,c), we use the participation number $N$,
\begin{equation}
N=\sum_{x,y}(|a_{x,y}|^4 + |b_{x,y}|^4),
\end{equation}
which measures the degree of localization of the mode, and the modal frequency $\omega$. \revision{The frequency can be either in the bulk band gap (when it exists), or overlapping with bulk modes, which helps to distinguish between bulk and edge lasing in the small, experimentally-feasible system we consider, where the distinction between bulk, edge, and corner modes may not be sharp.}

\begin{figure}
\centering\includegraphics[width=\columnwidth]{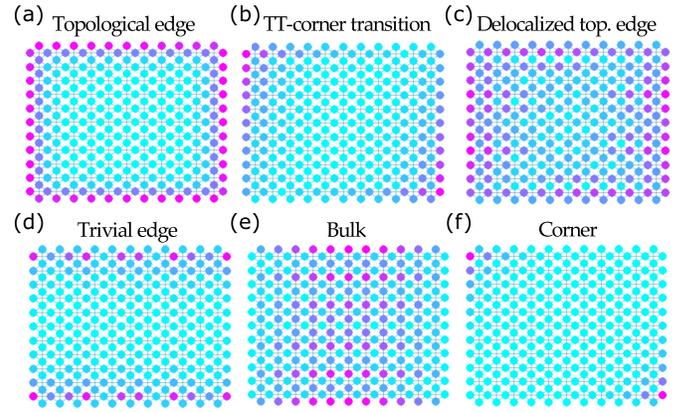}
\caption{Field intensity profiles of the dominant modes of the edge-pumped $C_4$-symmetric lattice in the different lasing regimes indicated in Fig.~\ref{fig:fig2}.}
\label{fig:edgetobulk}
\end{figure}

For small pump rates and detuning, topological edge modes have the highest gain [region marked TS in Fig.~\ref{fig:fig2}(b,c)].  A representative mode profile in this region is shown in Fig. \ref{fig:edgetobulk}(a). As the detuning increases, these modes gradually delocalize and penetrate into the bulk [TD, Fig.~\ref{fig:edgetobulk}(c)], corresponding to a low participation number. Interestingly, this delocalization starts to occur even before the bulk topological transition at $M=2$, due to the lasing modes' frequency approaching the bulk band edge, \revision{which results in a continuous increase of $N$}.

\revision{For small pump rates, above $M=2$ (just within the trivial phase) non-topological edge modes lase first [NE, Fig.~\ref{fig:edgetobulk}(d)]. These trivial edge states can be regarded as remnants of the topological ones, which emerge because the edge itself acts as defect: bulk sites are each coupled to 6 neighbours, while edge sites are only coupled to 3 neighbours. The trivial states are localized to either the horizontal or vertical edges and are unable to circulate around the entire edge.} At even higher detunings bulk modes become dominant [Fig.~\ref{fig:edgetobulk}(e)].

When both detuning and pump rate are high, the dominant modes are confined to the corners, corresponding to the maximal participation number [Fig.~\ref{fig:edgetobulk}(f)]. \revision{These corner states are localized by the gain, with the detuning $M$ primarily affecting their frequency. Note that although Figs.~\ref{fig:edgetobulk}(b) and (f) exhibit qualitatively similar modal profiles, the modes have different frequencies; the former corresponds to a topological edge mode residing in the bulk band gap, whereas the latter has a frequency overlapping with one of the bulk bands. In addition, the participation number $N$ of the topological edge states increases with the gain/loss contrast $\gamma$, whereas $N$ saturates to a $\gamma$-independent value for the corner modes.}

Before merging into bulk modes, the dominant topological edge modes also exhibit frequency bifurcations. This bifurcated region is shown with label TT in Figs.~\ref{fig:fig2}(b,c). In this regime, two topological edge modes share the same maximum gain. Fig.~\ref{fig:fig2}(d) shows the frequency of the dominant mode as a function of detuning for $P=0.5J$ and $2J$. The dominant edge mode at low detuning has zero frequency, while in the bifurcated regime, the frequency splits. Interestingly, this bifurcation occurs close to the critical detuning where the bulk band gap is maximized, corresponding to the strongest robustness of the edge modes to disorder. Evidently, this protection against disorder does not translate to protection of single edge mode lasing. For moderate gain $P_{edge}=0.5J$, above $M=2J$, the transition to the bulk mode lasing is also apparent as a discontinuity in the lasing frequency as a function of $M$. 

We can understand this bifurcation by inspecting the lasing mode profiles in Fig.~\ref{fig:edgetobulk}. Before the bifurcation the modal profile shown in Fig.~\ref{fig:edgetobulk}(a) resides on both sublattices, experiencing net gain along the entire edge. Small detunings $M$ redistribute the intensity between the sublattices while preserving the mid-gap lasing frequency $\omega = 0$. After the bifurcation, the lasing modes become asymmetric, see e.g. Fig.~\ref{fig:edgetobulk}(b). The positive $\omega$ mode preferentially excites the $a$ sublattice, resulting in net gain while propagating along the top and bottom edges and net loss along the left and right edges. The negative $\omega$ mode, preferentially localized to the $b$ sublattice, exhibits the opposite behaviour. With increasing pump rate and detuning these profiles continuously evolve either into corner-like modes similar to Fig.~\ref{fig:edgetobulk}(b) or they delocalize into the bulk.

%Fig. \ref{fig:edgetobulk} shows the representative profiles of the dominant modes. Topological edge modes circle the whole perimeter of the lattice. The bifurcated edge modes (a), continuously become corner-like or bulk-like, and mode profiles are similar to the transition regime (b). Just before the band-gap closes ($M\approx 2$, gain $\approx0$), the dominant modes are delocalized (c). The non-topological edge modes are confined to the edges, but they do not cover the whole perimeter (d). At high detuning but small pump rates, bulk modes dominate (e). And at large detuning and large pump rates, corner modes are the dominant (f). 

\begin{figure}
\centering\includegraphics[width=\columnwidth]{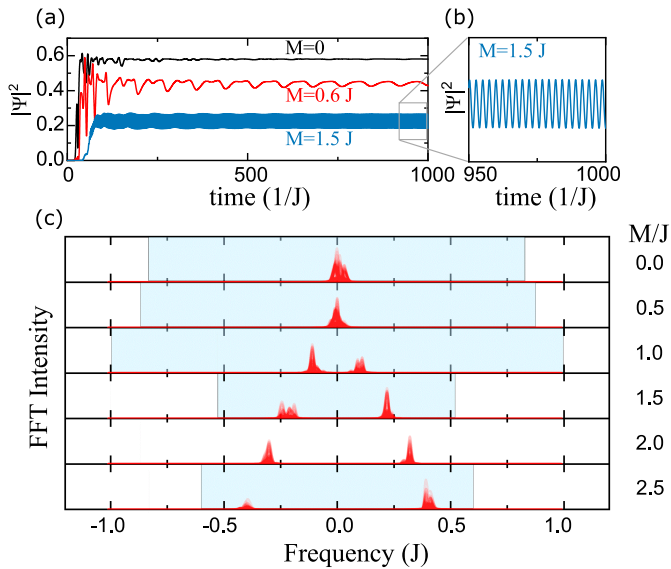}
\caption{(a) Intensity at a selected edge site as a function of time for detunings $M=0, 0.6$ and $1.5$. (b) Zoomed-in intensity of the $M=1.5$ case. (c) Intensity of the Fourier transform of the field amplitude at each lattice site, for various detunings. The bulk band gap is shaded in blue. We have used an edge pump rate $P_{edge} = J$ and uniform losses $\gamma=0.5J$.}
\label{fig:fft}
\end{figure}

To verify the stability of the dominant lasing modes and the ability to observe these bifurcations in experiment, we perform numerical simulations of the full nonlinear equations of motion~\eqref{eq:classA} using the 4th order Runge-Kutta method, initializing the field at time $t=0$ with random noise. Fig.~\ref{fig:fft}(a,b) shows the initial dynamics of the optical intensity at one edge resonator, while Fig.~\ref{fig:fft}(c) plots the field's frequency spectrum in the time interval $[100/J,200/J]$, after the initial transients have diminished. At low detuning the intensity reaches a steady state, with a single peak in its frequency spectrum centred at zero detuning. Approaching the onset of the bifurcation ($M=0.6J)$, relaxation times get longer due to many modes sharing the same gain, as shown in Fig. \ref{fig:fft}(a). Beyond this detuning there is indeed a splitting of the frequency spectrum into two distinct peaks, resulting in persistent intensity oscillations, Fig. \ref{fig:fft}(b).  These observations are all consistent with the above analysis of the linear spectrum. 

There are various ways one could suppress this bifurcation and preserve single edge mode lasing. First, the bifurcation emerges as a transition from net gain along the entire edge, to net gain only in localized regions. Therefore, one of the modes can be suppressed by pumping only one pair of edges, at the expense of increasing the lasing threshold. Alternatively, one could consider spatial modulation of the sublattice detuning $M$, keeping $M$ small at the edge sites to maintain strong overlap between the edge mode and the pumped sublattice, and using larger values of $M$ in the bulk to maximize the band gap and more strongly localize the topological modes to the edge, reducing the lasing threshold. Finally, for sufficiently large detunings $M$ a narrow band gain medium may be sufficient to isolate one of the modes.

In summary, we have analyzed the lasing modes of a tunable two-dimensional topological resonator lattice, \revision{focusing on the effects} of a frequency detuning between its two sublattices and the pump strength on the lattice's dominant lasing modes, which can vary from edge modes to bulk and corner modes. \revision{We have found that the increasing the frequency detuning first induces a bifurcation of the topological edge states from single to two mode lasing, before eventually they shift into resonance with bulk modes and delocalize.} Thus, even in the topological phase, the topological edge state lasing can exhibit bifurcations between single mode lasing (at the middle of the band gap) and two mode lasing (at frequencies symmetric about the gap center), resulting from the interplay between the sublattice detuning and the pumping profile. Simulations of a class A laser model close to its lasing threshold are consistent with our analysis of the system's linear modes.

\paragraph{\large{Funding.}}Institute for Basic Science (IBS-R024-Y1, IBS-R024-Y2).

\paragraph{\large{Disclosure.}}The authors declare no conflicts of interest

% Bibliography
\bibliography{sample}

\begin{thebibliography}{10}
\newcommand{\enquote}[1]{``#1''}

\bibitem{Ozawa2019}
T.~Ozawa, H.~M. Price, A.~Amo, N.~Goldman, M.~Hafezi, L.~Lu, M.~C. Rechtsman,
  D.~Schuster, J.~Simon, O.~Zilberberg, and I.~Carusotto, \enquote{Topological
  photonics,} {\protect\JournalTitle{Reviews of Modern Physics}} \textbf{91}
  (2019).

\bibitem{smirnova2019nonlinear}
D.~Smirnova, D.~Leykam, Y.~Chong, and Y.~Kivshar, \enquote{Nonlinear
  topological photonics,} {\protect\JournalTitle{Applied Physics Reviews}}
  \textbf{7}, 021306 (2020).

\bibitem{Ota_2020}
Y.~Ota, K.~Takata, T.~Ozawa, A.~Amo, Z.~Jia, B.~Kante, M.~Notomi, Y.~Arakawa,
  and S.~Iwamoto, \enquote{Active topological photonics,}
  {\protect\JournalTitle{Nanophotonics}} \textbf{9}, 547–567 (2020).

\bibitem{Feng2017}
L.~Feng, R.~El-Ganainy, and L.~Ge, \enquote{Non-hermitian photonics based on
  parity-time symmetry,} {\protect\JournalTitle{Nature Photonics}} \textbf{11},
  752--762 (2017).

\bibitem{Bahari2017}
B.~Bahari, A.~Ndao, F.~Vallini, A.~E. Amili, Y.~Fainman, and B.~Kant{\'{e}},
  \enquote{Nonreciprocal lasing in topological cavities of arbitrary
  geometries,} {\protect\JournalTitle{Science}} \textbf{358}, 636--640 (2017).

\bibitem{Ota2018}
Y.~Ota, R.~Katsumi, K.~Watanabe, S.~Iwamoto, and Y.~Arakawa,
  \enquote{Topological photonic crystal nanocavity laser,}
  {\protect\JournalTitle{Communications Physics}} \textbf{1}, 86 (2018).

\bibitem{Han2019}
C.~Han, M.~Lee, S.~Callard, C.~Seassal, and H.~Jeon, \enquote{Lasing at
  topological edge states in a photonic crystal {L3} nanocavity dimer array,}
  {\protect\JournalTitle{Light: Science {\&} Applications}} \textbf{8} (2019).

\bibitem{Zeng2020}
Y.~Zeng, U.~Chattopadhyay, B.~Zhu, B.~Qiang, J.~Li, Y.~Jin, L.~Li, A.~G.
  Davies, E.~H. Linfield, B.~Zhang, Y.~Chong, and Q.~J. Wang,
  \enquote{Electrically pumped topological laser with valley edge modes,}
  {\protect\JournalTitle{Nature}} \textbf{578}, 246--250 (2020).

\bibitem{StJean2017}
P.~St-Jean, V.~Goblot, E.~Galopin, A.~Lema{\^{\i}}tre, T.~Ozawa, L.~L. Gratiet,
  I.~Sagnes, J.~Bloch, and A.~Amo, \enquote{Lasing in topological edge states
  of a one-dimensional lattice,} {\protect\JournalTitle{Nature Photonics}}
  \textbf{11}, 651--656 (2017).

\bibitem{Parto2018}
M.~Parto, S.~Wittek, H.~Hodaei, G.~Harari, M.~A. Bandres, J.~Ren, M.~C.
  Rechtsman, M.~Segev, D.~N. Christodoulides, and M.~Khajavikhan,
  \enquote{Edge-mode lasing in 1d topological active arrays,}
  {\protect\JournalTitle{Physical Review Letters}} \textbf{120} (2018).

\bibitem{Zhao2018}
H.~Zhao, P.~Miao, M.~H. Teimourpour, S.~Malzard, R.~El-Ganainy, H.~Schomerus,
  and L.~Feng, \enquote{Topological hybrid silicon microlasers,}
  {\protect\JournalTitle{Nature Communications}} \textbf{9} (2018).

\bibitem{Harari2018}
G.~Harari, M.~A. Bandres, Y.~Lumer, M.~C. Rechtsman, Y.~D. Chong,
  M.~Khajavikhan, D.~N. Christodoulides, and M.~Segev, \enquote{Topological
  insulator laser: Theory,} {\protect\JournalTitle{Science}} \textbf{359},
  eaar4003 (2018).

\bibitem{Bandres2018}
M.~A. Bandres, S.~Wittek, G.~Harari, M.~Parto, J.~Ren, M.~Segev, D.~N.
  Christodoulides, and M.~Khajavikhan, \enquote{Topological insulator laser:
  Experiments,} {\protect\JournalTitle{Science}} \textbf{359}, eaar4005 (2018).

\bibitem{Schomerus2013}
H.~Schomerus, \enquote{Topologically protected midgap states in complex
  photonic lattices,} {\protect\JournalTitle{Optics Letters}} \textbf{38}, 1912
  (2013).

\bibitem{Weimann2016}
S.~Weimann, M.~Kremer, Y.~Plotnik, Y.~Lumer, S.~Nolte, K.~G. Makris, M.~Segev,
  M.~C. Rechtsman, and A.~Szameit, \enquote{Topologically protected bound
  states in photonic parity{\textendash}time-symmetric crystals,}
  {\protect\JournalTitle{Nature Materials}} \textbf{16}, 433--438 (2016).

\bibitem{Hafezi2011}
M.~Hafezi, E.~A. Demler, M.~D. Lukin, and J.~M. Taylor, \enquote{Robust optical
  delay lines with topological protection,} {\protect\JournalTitle{Nature
  Physics}} \textbf{7}, 907--912 (2011).

\bibitem{Secl2019}
M.~Secl{\`{\i}}, M.~Capone, and I.~Carusotto, \enquote{Theory of chiral edge
  state lasing in a two-dimensional topological system,}
  {\protect\JournalTitle{Physical Review Research}} \textbf{1} (2019).

\bibitem{Amelio2019}
I.~Amelio and I.~Carusotto, \enquote{Theory of the coherence of topological
  lasers,} {\protect\JournalTitle{arXiv eprints}} p. arXiv:1911.10437 (2019).

\bibitem{Longhi2018}
S.~Longhi, Y.~Kominis, and V.~Kovanis, \enquote{Presence of temporal dynamical
  instabilities in topological insulator lasers,} {\protect\JournalTitle{{EPL}
  (Europhysics Letters)}} \textbf{122}, 14004 (2018).

\bibitem{Leykam2018}
D.~Leykam, S.~Mittal, M.~Hafezi, and Y.~Chong, \enquote{Reconfigurable
  topological phases in next-nearest-neighbor coupled resonator lattices,}
  {\protect\JournalTitle{Physical Review Letters}} \textbf{121}, 023901 (2018).

\bibitem{Mittal2019}
S.~Mittal, V.~V. Orre, D.~Leykam, Y.~D. Chong, and M.~Hafezi, \enquote{Photonic
  anomalous quantum {H}all effect,} {\protect\JournalTitle{Phys. Rev. Lett.}}
  \textbf{123}, 043201 (2019).

\end{thebibliography}

% Full bibliography added automatically for Optics Letters submissions; the following line will simply be ignored if submitting to other journals.
% Note that this extra page will not count against page length
\bibliographyfullrefs{sample}

%Manual citation list
%\begin{thebibliography}{1}
%\bibitem{Zhang:14}
%Y.~Zhang, S.~Qiao, L.~Sun, Q.~W. Shi, W.~Huang, %L.~Li, and Z.~Yang,
 % \enquote{Photoinduced active terahertz metamaterials with nanostructured
  %vanadium dioxide film deposited by sol-gel method,} Opt. Express \textbf{22},
  %11070--11078 (2014).
%\end{thebibliography}

% Please include bios and photos of all authors for aop articles
\ifthenelse{\equal{\journalref}{aop}}{%
\section*{Author Biographies}
\begingroup
\setlength\intextsep{0pt}
\begin{minipage}[t][6.3cm][t]{1.0\textwidth} % Adjust height [6.3cm] as required for separation of bio photos.
  \begin{wrapfigure}{L}{0.25\textwidth}
    \includegraphics[width=0.25\textwidth]{john_smith.eps}
  \end{wrapfigure}
  \noindent
  {\bfseries John Smith} received his BSc (Mathematics) in 2000 from The University of Maryland. His research interests include lasers and optics.
\end{minipage}
\begin{minipage}{1.0\textwidth}
  \begin{wrapfigure}{L}{0.25\textwidth}
    \includegraphics[width=0.25\textwidth]{alice_smith.eps}
  \end{wrapfigure}
  \noindent
  {\bfseries Alice Smith} also received her BSc (Mathematics) in 2000 from The University of Maryland. Her research interests also include lasers and optics.
\end{minipage}
\endgroup
}{}

\end{document}